\documentstyle[epsfig]{lamuphys}
\makeatletter
\let\chapter\hid@chapter
\makeatother
\newcommand{\be}{\begin{equation}}
\newcommand{\ee}{\end{equation}}
\newcommand{\ba}{\begin{eqnarray}}
\newcommand{\ea}{\end{eqnarray}}
\begin{document}
\pagenumbering{arabic}
\thispagestyle{empty}
\setcounter{page}{0}
\begin{flushright}
ZU-TH 34/97\\
November 1997
\end{flushright}

\vfill

\begin{center}
{\LARGE \bf The $\Delta$(1232) as an Effective Degree of \\[10pt]
Freedom in Chiral Perturbation Theory*}\\[40pt]

\large
Joachim Kambor\\[0.5cm]
Institut f\"ur Theoretische Physik, Universit\"at Z\"urich\\
CH-8057 Z\"urich, Switzerland \\[10pt]
\end{center}
\vfill

\begin{abstract}
Heavy baryon chiral perturbation theory including spin 3/2 delta 
resonances as effective degrees of freedom is reviewed. The theory 
admits a systematic expansion in the small scale $\epsilon$, 
where $\epsilon$ collectively denotes soft momenta, the pion mass or
the delta-nucleon mass difference. Renormalization is discussed in
some detail on the example of the scalar sector of one-nucleon
processes, and a reformulation of the principle of resonance
saturation for counterterms of the HBChPT lagrangian is sketched.
As an application, the polarizabilities of the nucleon are discussed 
at order $\epsilon^3$.
\end{abstract}

\vfill
\begin{center}
Talk given at the Workshop on Chiral Dynamics 1997 \\[4pt] 
Mainz, Germany, Sept. 1 - 5, 1997 \\[4pt]
To appear in the Proceedings
\end{center}

\vfill
\noindent * Work supported in part by Schweizerischer Nationalfonds.

\newpage

\title{The $\Delta$(1232) as an Effective Degree of Freedom in Chiral
Perturbation Theory}

\author{Joachim Kambor}

\institute{Institut f\"ur Theoretische Physik, Universit\"at Z\"urich,\\
CH-8057 Z\"urich, Switzerland}

\authorrunning{Joachim Kambor}
\titlerunning{The $\Delta$(1232) as an Effective Degree of Freedom in ChPT}
\maketitle

\begin{abstract}
Heavy baryon chiral perturbation theory including spin 3/2 delta 
resonances as effective degrees of freedom is reviewed. The theory 
admits a systematic expansion in the small scale $\epsilon$, 
where $\epsilon$ collectively denotes soft momenta, the pion mass or
the delta-nucleon mass difference. Renormalization is discussed in
some detail on the example of the scalar sector of one-nucleon
processes, and a reformulation of the principle of resonance
saturation for counterterms of the HBChPT lagrangian is sketched.
As an application, the polarizabilities of the nucleon are discussed 
at order $\epsilon^3$.
\end{abstract}
\section{Introduction}
Heavy baryon chiral perturbation theory (HBChPT), the effective low 
energy theory of QCD for $\pi-N$ interactions, is by now a well
developed field. At first sight there appears to be a problem in 
doing chiral perturbation theory for nucleons -- the nucleon mass is
large compared to soft momenta or $M_\pi$, and it does not vanish in 
the chiral limit. In the pioneering work of (\cite{GSS88}) it was noted
that the relativistic formulation adopted there does not allow for
a consistent low energy expansion. Soon after it was shown
(\cite{JM9192}) how the situation can be cured by going to the extreme
non-relativistic limit, and thereafter \cite{BKKM92} developed a
consistent power counting sheme, i.e. HBChPT. Renormalization at order
$p^3$ has been thoroughly discussed in (\cite{E94}). Also, extensive 
calculations 
have been done since then (\cite{BKMrev95}) , showing the 
power of the method in an impressive way. Very recently, the missing 
link between the T-matrix elements calculated in HBChPT and the fully 
relativistic S-matrix elements one is actually seeking, has been
provided (\cite{EM97}). Several aspects of HBChPT are reviewed in
these procedings. (\cite{Bernard},\cite{Gasser},\cite{Meissner},
\cite{vanKolck})
 
Despite this progress, some issues remain open and are widely
discussed. One of these questions is the way in which the baryonic 
resonances are treated. In HBChPT the only explicit degrees of freedom
are the nucleons and the Goldstone Bosons, {\it i.e.} the pions. 
Resonances are also included, but only in the form of local
counterterms. They are thought to contribute to the coupling constants
of these counterterms, suppressed by a heavy scale, and this is the
way in which most work to date has been done. 
\footnote{Exceptions include the original work of \cite{JM9192}, as
well as applications to the case of chiral SU(3)xSU(3), (\cite{BSS93}).}
However, the $\Delta$(1232) resonance has a special status in two respects.
First, it lies only about 300 MeV above the nucleon ground state --
treating it as a heavy state seems therefore to be of questionable 
validity from a phenomenological point of view. Also, it couples very 
strongly to the $\pi N$-system, and
quite generally contributes substantially through resonance
exchange graphs in those channels where such effects are possible.  
A second, more conceptually motivated criticism arises from the way 
the nucleons itself are treated: as heavy static sources. The
nonvanishing of the nucleon mass forces us into HBChPT in order to
maintain a consistent chiral power counting scheme. Technically
speaking we are expanding in powers of $1/m$, with $m$ beeing the nucleon mass 
in the chiral limit. The symmetry limit is thus obtained by sending $m$
to infinity. This consideration shows that the scale which suppresses 
baryonic resonance contributions to local counterterms in HBChPT in 
general
cannot be the resonance mass, but rather it must be related to the 
mass difference, $\Delta=M_{\rm res}-m$. In the case of the delta 
resonance, this mass difference is phenomenologically small. While
this is an indication that it might be dangerous to expand in
$1/\Delta$, it does not tell us whether $\Delta$ is a small scale 
intrinsically. In the large $N_c$ limit the nucleon and delta
resonances are degenerate indeed, and an interesting approach is therefore 
to consider 
a combined chiral and $1/N_c$ expansion. Attempts in this direction 
have been made in (\cite{LMR94}).

Here we are less ambitious and {\it assume} that the scale $\Delta$ is 
small compared to the scale of chiral symmetry breaking or the nucleon
mass. I describe a scheme by which the delta contributions can be
treated in a systematic power expansion in soft momenta, the pion 
mass and the mass difference $\Delta$, collectively denoted by
$\epsilon$, which has been developed recently (\cite{HHK97a}).
In section 2 the formalism of HBChPT is briefly reviewed and the 
steps necessary to include the delta degrees of freedom are sketched. 
In section 3 I discuss renormalization of this theory, using the
scalar sector of the $\pi N$-system as an example. We will compare
the theory with and without explicit $\Delta(1232)$ by studying the 
limit $M_\pi/\Delta \rightarrow 0$ in the former. A new  
formulation of resonance saturation with ``light deltas'' naturally
emerges and is sketched in
subsection 3.3. Section 4 discusses an application of the formalism
to the polarizabilities of the nucleon at order $\epsilon^3$
(\cite{HHK97b}), where 
interesting effects have been found. In the final section I draw the
conclusions and mention some directions for future work.

\section{Inclusion of $\Delta(1232)$: 1/m-expansion}

\subsection{HBChPT without $\Delta(1232)$}

I briefly review HBChPT and it's derivation using a
$1/m$-expansion. 
The subsection serves to set the notation, but also
to point out the relevance of $1/m$-corrections.

Consider the relativistic formulation of ChPT for the $\pi N$-system
and write the effective lagrangian as a string of terms (\cite{GSS88})
\begin{equation}
{\cal L}_{\pi N}={\cal L}_{\pi N}^{(1)}+{\cal L}_{\pi N}^{(2)}+ \ldots
\end{equation}
where the superscript denotes the number of derivatives. The first term in this
expansion reads 
\begin{equation}
{\cal L}_{\pi N}^{(1)}=\bar N \left( i\not\!\!{D}-\dot{m}+i{\dot{g}_A \over 2} 
\not\!{u} \gamma_5 \right) N
\label{LpiN1}
\end{equation}
where $N$ is the nucleon field, $\dot{m}$ and $\dot{g}_A$ are the nucleon 
mass and axial-vector coupling constant in the chiral limit, respectively.

In the chiral limit, $\dot{m}\not=0$. As a consequence, the covariant derivative
on $N$ counts as order one ($p$ denotes a generic soft momentum)
\begin{equation}
{D}_\mu N= O(1) \qquad{\rm but}\qquad
(i\not\!\!{D}-\dot{m}) N = O(p).
\end{equation}
Therefore, the loop-expansion is not equivalent to a low-energy expansion
(\cite{GSS88}), in contrast to the Goldstone boson sector (\cite{Wei79}).
The problem can be overcome by going to the extreme nonrelativistic limit
(\cite{JM9192}).
The idea is to move the $\dot{m}$-dependence from the propagator to the 
vertices. This can be achieved by choosing the frame dependent
decomposition with fixed four-velocity $v_\mu$
\begin{equation}
H_v = \exp\{i\dot{m} (v\cdot x)\} P_v^+ N , \qquad
h_v = \exp\{i\dot{m} (v\cdot x)\} P_v^- N  
\label{decomp}
\end{equation} 
with
\begin{equation}
P_v^\pm={1\pm \not\!{v} \over 2} \ .
\label{Ppm}
\end{equation}
Using a path integral formulation, the heavy degrees of freedom,
$h_v$, can be integrated out systematically (\cite{BKKM92}), leading
to the effective action
\begin{equation}
\hat S_{\pi N}=\int \bar H_v \left( {\cal A}+\gamma_0 {\cal B}^\dagger \gamma_0
{\cal C}^{-1} {\cal B}\right) H_v \ . 
\label{effaction}
\end{equation}
${\cal A,B,C}$ admit the low energy expansions
\begin{eqnarray}
{\cal A} &=& i (v\cdot D) +\dot{g}_A (S\cdot u)+\ldots  \nonumber \\
{\cal B} &=& i \not\!{\nabla}^\bot -{\dot{g}_A \over 2} v\cdot u
\gamma_5 +\ldots \\
{\cal C} &=& 2 \dot{m} +i (v\cdot D) +\dot{g}_A (S\cdot u)+\ldots  \nonumber \\
\label{ABC}
\end{eqnarray}
where $S_\mu$ denotes the Pauli-Lubanski spin vector, $u_\mu$
contains the pion field in the standard manner and the dots denote
terms of order $p^2$. Writing the effective 
lagrangian in terms of fields 
$H_v$ ensures a consistent low energy expansion. Decomposing the 
nucleon four-momentum according to $p_\mu=\dot{m} v_\mu+k_\mu$, 
where $k_\mu$ is a soft residual momentum, the propagator reads
\begin{equation}
S(\omega)={i \over v\cdot k +i\epsilon},
\end{equation}
with $\omega=v\cdot k$. The low energy expansion of the $\pi N$-system 
so obtained is a simultaneous expansion in 
\begin{equation}
{p\over 4\pi F_\pi} \qquad {\rm and} \qquad {p\over \dot{m}}\ . 
\end{equation}
The approach shows that the terms arising from the $1/m$-expansion, 
i.e. the second term on the right hand side of Eq. (\ref{effaction}),
have fixed coefficients given in terms of the coupling constants of the
relativistic chiral lagrangian. Otherwise the theory would not be Lorentz 
invariant (\cite{EM96}). The last observation can alternatively be derived 
by employing reparametrization invariance (\cite{LM92}). However, the
formalism employed here gives a physical interpretation of these
$1/m$-corrections: they arise from integrating out the heavy component
of the nucleon field. Renormalization at
the one-loop level (up to order $p^3$) is thoroughly discussed in
(\cite{E94}).
Moreover, it has been shown recently that the 
T-matrix elements calculated in HBChPT are not sufficient to recover
all relativistic S-matrix elements. (\cite{EM97}) It therefore seems
to be unavoidable to start from a relativistic formulation of the
$\pi N$-system. The subsequent choice of heavy baryon fields $H_v$ as
well as the $1/m$-expansion is only a vehicle to perform the
loop-expansion in a systematic manner.  

\subsection{Local versus non-local contributions of resonances}

In the introduction it was argued that the spin 3/2 delta resonances
play a special role in the $\pi N$-system. Before going
into the discussion of including $\Delta$(1232) in HBChPT, I will
show, on a simple example, why the treatment of the delta resonance as 
local counterterms could be problematic. (\cite{Kam96}) 

Consider the magnetic polarizability of the nucleon in HBChPT. 
It has a low energy expansion of 
the form (modulo logarithms of $M_\pi$) \cite{BKKM92}
\begin{equation}
\beta={{\rm const.} \over M_\pi} \left\{ 1+c_1 {M_\pi\over\Lambda}
+c_2 {M_\pi^2 \over \Lambda^2}+\ldots \right\}
\label{betaexp}
\end{equation}
where $\Lambda\in \left\{ 4\pi F_\pi, m_N\right\}$ is a heavy scale and 
the $c_i$ are dimensionless constants. This expansion is well suited to 
derive low-energy theorems (LET) (\cite{EMLet95}), valid in the chiral limit, 
i.e.
\begin{equation}
\lim_{M_\pi \rightarrow 0} M_\pi \beta= {\rm const.} 
\end{equation}
In the physical world of finite quark masses, however, the series 
(\ref{betaexp}) might converge slowly, due to large coefficients $c_i$.
Consider the effect of delta exchange on $\beta$, as shown in Fig. 1.
If we shrink the delta propagator to a point, the constants $c_i$ will
pick contributions of the form
\begin{equation}
c_i\sim \left( {m_\rho \over \Delta } \right)^i,
\label{ciestimate}
\end{equation}
where
\begin{equation}
\Delta\equiv m_\Delta-m_N|_{m_q\rightarrow 0} \approx 300\ {\rm MeV}
\end{equation}
is small and of the same order as a typical low energy scale, e.g. $M_\pi$. 
\begin{figure}[t]
\begin{center}
\leavevmode
\hbox{%
\epsfxsize=10.0truecm
\epsfbox{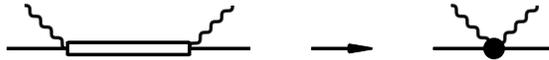} }
\end{center}
\caption{$\Delta$-resonance exchange contribution to nucleon Compton 
scattering. Single, double and wavy lines denote nucleons, delta and 
photons, respectively. The shaded dot denotes a local counterterm.}
\end{figure}
In Eq. (\ref{ciestimate}) I have inserted $m_\rho$ as a typical
hadronic scale. Then the individual terms in the bracket in (\ref{betaexp}) are
all of order one and it will be necessary to resum the series. The 
point is that the scale $\Delta$ appears in the denominator, not $m_\Delta$. 
The correct limit to be considered is $m_\Delta\rightarrow \infty$, 
$m_N\rightarrow \infty$, with $\Delta$ fixed.
We cannot treat the delta resonances as heavy compared to the 
nucleon a priori. The concept of resonance saturation is in this case at least 
questionable and it's applicability warrants further discussion.

\subsection{Inclusion of $\Delta$(1232) in HBChPT: Formalism}

The conceptual problems described are related to the fact that the
delta resonances (and also the nucleon!) contain both, heavy and light degrees
of freedom. The light components of the delta resonances decouple only in 
the strict chiral limit (\cite{JM9192}). Jenkins and Manohar have included 
the Baryon
decuplet as dynamical degree of freedom in the formulation of HBChPT right from
the beginning. Here I will describe work in collaboration 
with Hemmert and Holstein, where we show explicitly, 
by means of a systematic $1/m$-expansion, how the framework of
(\cite{JM9192}) 
can be obtained from the relativistic formulation of the 
$\pi N\Delta$-interactions.(\cite{HHK97a}) 
The aim is to construct a low energy effective theory of
pions, nucleons and delta resonances. 
\footnote{We work in chiral SU(2) in order to keep things as simple as
possible. Generalization to chiral SU(3) is in principle possible but
has not yet been worked out in detail.}
We therefore emphasize the importance of including all possible
counterterms allowed by chiral symmetry. The theory which emerges
(``small scale expansion'') admits a systematic simultaneous expansion in 
\begin{equation}
{M_\pi\over \Lambda}, \quad {p\over \Lambda} \qquad {\rm and} \qquad
{\Delta\over \Lambda}, \qquad {M_\pi \over \Delta}={\rm const.}
\end{equation} 

We now turn to some technical details concerning the construction of
this effective field theory. 
Consider the lagrangian for a relativistic spin 3/2 field $\Psi_\mu$ coupled 
in a chirally invariant manner to the Goldstone bosons
\footnote{For the sake of simplicity, we suppress isospin indices throughout.}
\begin{eqnarray}
{\cal L}_{3/2}&=&\bar{\psi}^\mu\Lambda_{\mu\nu}\psi^\nu\nonumber\\
\Lambda_{\mu \nu} &=& - \; [ ( i \not\!\!{D} - 
                              m_{\Delta} ) g_{\mu \nu} - \frac{1}{4} 
                              \gamma_{\mu} \gamma^{\lambda} ( i \not\!\!{D} - 
                              m_{\Delta} ) \gamma_{\lambda} \gamma_{\nu} 
                              \nonumber \\
                  && + \frac{g_{1}}{2} g_{\mu \nu} \not\!{u} 
                              \gamma_{5} + \frac{g_{2}}{2} ( \gamma_{\mu} 
                              u_{\nu} + u_{\mu} \gamma_{\nu} ) \gamma_{5} 
                      + \frac{g_{3}}{2} \gamma_{\mu} \not\!{u} 
                              \gamma_{5} \gamma_{\nu} ]. 
\label{eq:Lambda}
\end{eqnarray}
We have factored out the dependence on the unphysical free parameter A using
$\psi_\mu(x)=(g_{\mu\nu}+ A/2 \gamma_\mu \gamma_\nu)\Psi^\nu(x)$ 
(\cite{Pas94}). 
The pion fields are contained in 
$u= \exp \left(i\vec{\tau} \cdot \vec{\pi}/F_\pi \right)$
and $D_\mu\psi_\nu$ denotes the covariant derivative 
on the spin 3/2 field.
The first two pieces in (\ref{eq:Lambda}) are the kinetic and
mass terms of a free Rarita-Schwinger Spinor (\cite{RS41},\cite{BDM89}). 
The remaining terms
constitute the most general chiral invariant couplings to pions. 

The next step consists of identifying the light and heavy degrees of freedom
of the spin 3/2 fields, respectively. The procedure is analogous to the case 
of spin 1/2 fields (\cite{BKKM92}), but technically more complicated 
due to the off-shell spin 1/2 degrees of freedom of the Rarita-Schwinger
field. In order to get rid of the mass dependence in (\ref{eq:Lambda}) 
we introduce the spin 3/2 projection operator for fields with {\it 
fixed velocity} $v_\mu$
\begin{equation}
P^{3/2}_{(33)\mu \nu} =  g_{\mu \nu} - \frac{1}{3} \gamma_{\mu} \gamma_{
                               \nu} - \frac{1}{3} \left( \not\!{v} \gamma_{\mu} 
                               v_{\nu} + v_{\mu} \gamma_{\nu}\not\!{v} \right) 
\label{eq:project}
\end{equation}
and identify the light degrees of freedom via
\begin{equation}
T_{\mu,v} (x) \equiv  P_{v}^{+} \; P^{3/2}_{(33)\mu\nu} \psi^{\nu}(x) 
               \; \exp (i m v \cdot x) . 
\label{eq:T}
\end{equation}
These fields coincide with the decuplet fields used in (\cite{JM9192}) and 
satisfy $v_\mu T^\mu_v=\gamma_\mu T^\mu_v=0$.
The remaining components, denoted collectively by $G_{\mu, v}$, are heavy 
and will be integrated out.
However, the effects of these degrees of freedom are 
included as they will give rise to $1/m$-corrections.

We now perform a systematic $1/m$-expansion, in analogy to the heavy 
nucleon formalism.  
We write the most general lagrangian as (suppressing all indices)
\begin{equation}
{\cal L}={\cal L}_{N} + {\cal L}_{\Delta} + {\cal L}_{\Delta N}
\end{equation}
with
\begin{eqnarray}
{\cal L}_{N} &=& \bar{H} {\cal A}_{N} H + ( \bar{h} {\cal B}_{N} H + h.c. ) + 
                  \bar{h} {\cal C}_{N} h \nonumber \\
{\cal L}_{\Delta N} &=& \bar{T} {\cal A}_{\Delta N} H + \bar{G} 
{\cal B}_{\Delta N} H + 
                         \bar{h} {\cal B}_{N \Delta} T + \bar G 
{\cal C}_{\Delta N} h +
                         h.c. \nonumber\\
{\cal L}_{\Delta}&=&\bar{T} {\cal A}_{\Delta} T + \bar{G} 
              {\cal B}_{\Delta} T + \bar{T} \gamma_{0} {\cal 
              B}_{\Delta}^{\dagger} \gamma_{0} G + \bar{G} {\cal C}_{\Delta} G.
\label{Lgeneral}
\end{eqnarray}

The matrices ${\cal A}_{N}$, ${\cal B}_{N}$, ...,${\cal C}_{\Delta}$ 
in (\ref{Lgeneral}) admit an expansion of the form
\begin{equation}
{\cal A}_\Delta = {\cal A}_\Delta^{(1)}+{\cal A}_\Delta^{(2)}+ ... ,
\end{equation}
where ${\cal A}_\Delta^{(n)}$ is of order $\epsilon^n$. 
Explicitly, the leading order contribution to 
${\cal A}_\Delta$ is
\begin{equation}
{\cal A}_{\Delta, \mu\nu}^{(1)} =
- \left[ i (v \cdot D)-\Delta + g_{1} (S \cdot u) \right] \; g_{\mu \nu}.
\label{llead}
\end{equation}
The heavy baryon propagator of the delta is thus 
$-i P^{3/2}_{(33)\mu\nu}/((v\cdot k)-\Delta)$ 
and hence counts as order $\epsilon^{-1}$ in our
expansion. Explicit expressions for the expansions of
${\cal B}_\Delta$, ${\cal C}_\Delta$ etc. can be found in (\cite{HHK97c}). 
Note that matrices ${\cal C}_\Delta$ and ${\cal C}_N$ start at order 
$\epsilon^0$.

The final step is again in analogy to the case where only nucleons were 
considered. Shifting variables and completing the square we obtain
the effective action
\begin{equation}
S_{\rm eff}= \int d^4x \left\{ \bar T \tilde {\cal A}_{\Delta} T
+\bar H \tilde {\cal A}_{N} H
+\left[ \bar T \tilde {\cal A}_{\Delta N} H + h.c.\right] \right\}
\label{Seff}
\end{equation}
with (I keep only leading order terms in $1/m$ here, for the sake of simplicity)
\begin{eqnarray}
\tilde {\cal A}_\Delta &=& {\cal A}_\Delta 
-\gamma_0  {\cal B}_{N \Delta}^\dagger \gamma_0 {\cal C}_N^{-1} 
{\cal B}_{N \Delta}
-\gamma_0 {\cal B}_\Delta^\dagger \gamma_0 {\cal C}_\Delta^{-1} {\cal B}_\Delta
\nonumber\\
\tilde {\cal A}_N &=& {\cal A}_N 
-\gamma_0  {\cal B}_{N}^\dagger \gamma_0 {\cal C}_N^{-1} 
{\cal B}_{N}
-\gamma_0 {\cal B}_{\Delta N}^\dagger \gamma_0 {\cal C}_\Delta^{-1} 
{\cal B}_{\Delta N}
\nonumber \\
\tilde {\cal A}_{\Delta N} &=& {\cal A}_{\Delta N} 
-\gamma_0  {\cal B}_{N \Delta}^\dagger \gamma_0  {\cal C}_N^{-1} 
{\cal B}_{N} 
-\gamma_0 {\cal B}_\Delta^\dagger \gamma_0 {\cal C}_\Delta^{-1} 
{\cal B}_{\Delta N} .
\label{Atilde}
\end{eqnarray}
The new terms in proportion to ${\cal C}_\Delta^{-1}$ and ${\cal C}_N^{-1}$
are given entirely in terms of coupling constants of the lagrangian for 
relativistic fields. This guarantees reparameterization
invariance (\cite{LM92}) and Lorentz invariance (\cite{EM96}). 
Also, these terms 
are $1/m$ suppressed. The effects of the heavy degrees of freedom 
(both spin 3/2 and 1/2) thus show up only at order $\epsilon^2$. Note also that 
the effective $NN$-, $N\Delta$- and $\Delta\Delta$-interactions all contain
contributions from both heavy $N$- and $\Delta$-exchange respectively.  

In the above formalism, it is understood that one has to include also the most 
general counterterm lagrangian consistent with chiral symmetry, Lorentz 
invariance, and the discrete symmetries P and C, in relativistic formulation.
The construction yields then 
automatically the contributions to matrices ${\cal A}$, 
${\cal B}$, and ${\cal C}$. 
In order to calculate a given process to order 
$\epsilon^n$, it then suffices to construct matrices ${\cal A}$ to the same 
order, $\epsilon^n$,
${\cal B}$ to order $\epsilon^{n-1}$, and ${\cal C}$ to order $\epsilon^{n-2}$. 
Finally one has to add all loop-graphs contributing at the order one is
working. The relevant diagrams can be found by straightforward power 
counting in $\epsilon$.

\section{The scalar sector to ${\cal O}(\epsilon^3)$: renormalization
and resonance saturation}

In this section we discuss renormalization on the example of the
scalar sector of the $\pi N$-system to order $\epsilon^3$. The aim is
to illustrate some important features of the formalism. First, the 
low energy constants have different meaning in the theory with 
and without explicit delta degrees of freedom. For instance, in the 
small scale expansion, the bare coupling constants have to absorb 
divergencies in proportion to $\Delta^3$, i.e. terms which do not
vanish in the chiral limit. We will explain why this is still consistent.
Second, the examples considered  
yield interesting phenomenological insights into the effects the
delta resonances have on low energy observables. By taking the limit 
$M_\pi/\Delta\rightarrow 0$ we can
study the convergence properties of these effects. Also, a natural and
systematic formulation of resonance saturation for ``light deltas''
emerges.

\subsection{Nucleon mass}

Starting from 
(\ref{Seff}) it is easily shown that, compared to the order $q^3$ calculation
in conventional HBChPT,  the only new contribution to the selfenergy at
order $\epsilon^3$ is due
to the diagram shown in Fig. 2 a), where the $\Delta$-propagator denotes the 
propagation of the light degrees of freedom $T_{\mu,v}$. 
The explicit result has been given in 
(\cite{BKM93}), our interpretation, however, is different. Thus, we
obtain for the nucleon mass
\begin{equation}
m_N={m}^r-4 c_1^r M_\pi^2-{3 {g}_A^2 M_\pi^3 \over 32 \pi F_\pi^2}
+{g_{\pi N\Delta}^2 M_\pi^3 \over 12 \pi^2 F_\pi^2} 
R\left({\Delta \over M_\pi}\right)
\label{mN}
\end{equation}
where 
\begin{eqnarray}
R(x)&=&-4(x^2-1)^{3/2} \ln\left( x +\sqrt{ x^2-1 } \right) \nonumber\\
&&+4 x^3 \ln {2 x} 
-x \left( 1+6 \ln {2 x}\right) .
\label{defR}
\end{eqnarray}
In (\ref{mN}) terms in proportion to $\Delta^3$ and 
$\Delta M_\pi^2$ have been absorbed in the renormalized coupling
constants ${m}^r$ and $c_1^r$, respectively,
\begin{equation}
{m}^r={m}+{g_{\pi N\Delta}^2 \over 3 F^2} \Delta^3
\left\{ -16 L+{1\over \pi^2} \left[{1 \over 3}
-\ln {2\Delta\over \mu} \right] \right\}
\label{mren}
\end{equation}
and 
\begin{equation}
c_1^r=c_1+{g_{\pi N\Delta}^2 \over 8 F^2} \Delta
\left\{ -16 L+{1\over \pi^2} \left[ -\ln {2\Delta\over \mu}
\right] \right\} .
\label{c1ren}
\end{equation}
$c_1$ is the coupling constant of a ${\cal O}(p^2)$ term in the chiral
lagrangian of the NN-sector, i.e.
\begin{equation}
\hat{{\cal L}}_{\pi N}^{(2)}=c_1 \bar H_v tr (\chi_+) H_v +\ldots,
\end{equation}
with
\begin{equation}
\chi_+= (u^\dagger \chi u^\dagger+u\chi^\dagger u), \qquad 
\chi=2 B_0 (s+i p),
\end{equation}
and $L$ contains the pole at d=4 in dimensional regularization
\begin{equation}
L={\mu^{d-4}\over 16 \pi^2} \left[ {1\over d-4}+
{1\over 2} (\gamma_E-1-\ln 4\pi)\right].
\label{defL}
\end{equation}
The bare nucleon mass parameter, $m$, and the bare coupling
constant $c_1$ are thus infintely renormalized. The
renormalized coupling constants do not depend on the quark masses,
however, and this is an important feature in order to show decoupling
of the delta degrees of freedom in the chiral limit.
\begin{figure}[t]
\begin{center}
\leavevmode
\hbox{%
\epsfxsize=10.0truecm
\epsfbox{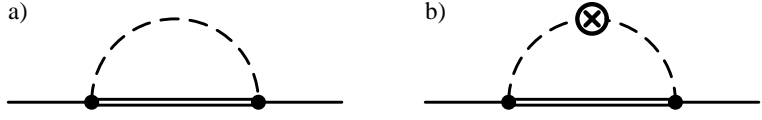} }
\end{center}
\caption{$\Delta$-loop contributions to a) nucleon selfenergy and
b) scalar form factor.
Single, double and dashed lines denote nucleons, delta and 
pions, respectively. The circled cross denotes the insertion of the 
scalar density external source.}
\end{figure}
Decoupling can be exemplified by taking the formal limit
$M_\pi/\Delta\rightarrow 0$, thereby expanding the function $R$ in
Eq. (\ref{defR}) according to
\begin{equation}
R\left({\Delta\over M_\pi}\right)=-{9 \over 8} {M_\pi\over \Delta} 
\left( 1+{4\over 3}\ln {2\Delta\over M_\pi} \right) 
+{\cal O}\left({M_\pi^3\over \Delta^3}\right).
\label{Rexp}
\end{equation}
This expansion shows explicitly that the ``light components'' of the
delta resonances start to contribute to the nucleon mass only at order
$p^4$, except for renormalization of the bare coupling constants $m$
and $c_1$. 

The fact that $m$ and $c_1$ are infinetely renormalized seems to be
puzzling at first sight. It is counterintuitive to the 
experience we have with 
renormalization in ChPT. For instance, one might think that $m$ is the
nucleon mass in the chiral limit, and thus it should stay finite. The
point is that we are not allowed to identify coupling constants of the
theory including delta degrees of freedom with those in HBChPT,
eventhough they multiply the same structures in the effective
lagrangians. The process of integrating out the additional degrees of
freedom leads to a (in general infinite) renormalization of the
bare coupling constants of the underlying theory.
A well known example of this type is given by the
comparison of the Goldstone Boson sector of chiral SU(3) with that of 
chiral SU(2). There, for instance, the pion decay constant in the
chiral SU(2) limit, $F$, is related to the decay constant in the
chiral SU(3) limit, $F_0$, via (\cite{GL85})
\begin{equation}
F=F_0 \left\{1+{\bar M_K^2 \over F_0^2} \left[
-{1 \over 32 \pi^2} \ln {\bar M_K^2 \over \mu^2}+8 L_4^r(\mu)
\right]\right\}.
\label{Frelation}
\end{equation}
Eq. (\ref{Frelation}) is valid to one-loop order. It is derived by 
calculating the mass corrections to $F_\pi$ in chiral SU(3) and then
taking the limit $\hat m \rightarrow 0$, with $m_s$ fixed. 
In the case discussed here, the situation is similar. The relevant
limit is the one taken in (\ref{Rexp}), i.e. $\hat m \rightarrow 0$,
but $\Delta$ fixed. Consequently, it is $m^r$ in (\ref{mren}) which is
to be identified with nucleon mass in the chiral limit.

Turning now to phenomenological consequences, we consider the 
function $R$ for physical masses 
$M_\pi=140\ {\rm MeV}$ and $\Delta\approx 2 M_\pi$. Each individual term in 
(\ref{defR}) is of order $\epsilon^0$, and $R$ suffers from large 
cancellations. Numerically we find 
$R(\Delta/M_\pi)=-1.60$. Using furthermore $g_{\pi N\Delta}=1.05 \pm
0.02$,
\footnote{This value is obtained from the $\Delta(1232)$ decay width
by employing the small scale expansion to leading order. It is more
consistent to use this value than the larger coupling obtained in
a relativistic Born model and employed for instance in (\cite{HHK97b}).}
this corresponds to a nucleon mass 
shift of about $-6\ {\rm MeV}$, somewhat smaller but of the same order
of magnitude than 
the $-15\ {\rm MeV}$ shift coming from the 
term in proportion to $M_\pi^3$. We can also study the ``convergence'' of the 
chiral expansion with respect to delta resonance effects by taking
into account only the leading term of order 
$M_\pi/\Delta$ in (\ref{Rexp}). This approximation yields again 
$R\approx -1.60$, i.e. the higher order terms in R can safely be 
neglected. We conclude that, for the nucleon mass, it is sufficient to work
to order $p^4$ in the chiral expansion. Note, however, that the 
small scale expansion presented here treats effects of comparable size,
$M_\pi^3$ and $M_\pi^4/\Delta$, at the same order of the expansion, 
viz. $\epsilon^3$. We expect this to improve the convergence 
of the perturbation series. 

\subsection{Scalar form factor and nucleon sigma term}

A similar analysis applies to the nucleon sigma term. The scalar form
factor of the nucleon is defined as
\begin{equation}
\sigma(t)=< N,p'| \hat m (\bar u u+\bar d d) |N,p>, \quad t=(p'-p)^2
\label{sigmadef}
\end{equation}
where $\hat m$ is the mean value of u and d quark mass. As for the
selfenergy, at ${\cal O}(\epsilon^3)$ there is only one additional 
diagram to be calculated compared to a $p^3$ HBChPT calculation,
\footnote{We are working in standard HChPT. The so called Generalized 
framework for HBChPT is considered in (\cite{BK97}).}
which
is shown in Fig. 2 b). Evaluating the result at $t=0$ we obtain the 
nucleon sigma term at order $\epsilon^3$
\begin{equation}
\sigma(0)=-4 c_1^r M_\pi^2-{9 \dot{g}_A^2 \over 64 \pi F_\pi^2}
M_\pi^3 + {g_{\pi N\Delta}^2 \over 4 \pi^2 F_\pi^2} M_\pi^3
S\left({\Delta\over M_\pi}\right),
\label{sigma0}
\end{equation}
where the function $S$ is defined as
\begin{equation}
S(x)=2 (x-1)^{1/2} \ln \left( x+ \sqrt{x^2-1} \right)-2 x \ln (2 x) .
\label{Sdef}
\end{equation}
$S$ is related to $R$ in Eq. (\ref{defR}) via the Feynman-Hellman theorem.
The expansion around the chiral limit reads
\begin{equation}
S\left( {\Delta\over M_\pi}\right)=
- {M_\pi \over\Delta} \left( {1\over 2} +\ln {2 \Delta \over M_\pi}
\right) + {\cal O}\left( M_\pi^3 \over \Delta^3\right) .
\label{Sexp}
\end{equation}
We observe that the delta-loop effects in the nucleon sigma term show
up at order $M_\pi^4$, as required by decoupling. 

Phenomenologically, the leading order term in the expansion of $S$
gives almost the full result. Taking input parameters $M_\pi=140$ MeV
and $\Delta\approx 2 M_\pi$ we obtain
\begin{equation}
S\left({\Delta\over M_\pi}\right)=-0.98, \qquad\qquad 
S\left({\Delta\over M_\pi}\right)|_{M_\pi^4}=-0.94.
\end{equation}
As in the case of the nucleon mass, it is sufficient to work to order 
$p^4$ in the chiral expansion in order to get the leading effect of 
intermediate ``light deltas''.

We can now solve for the renormalized coupling constant $c_1^r$, yielding
\begin{equation}
c_1^r=-{1\over 4 M_\pi^2}\left[
\sigma(0) +{9 \dot{g}_A^2 \over 64 \pi F^2} M_\pi^3 
-{g_{\pi N\Delta}^2 \over 4\pi^2 F^2} M_\pi^3 
S\left({\Delta\over M_\pi}\right)\right].
\label{c1renfix}
\end{equation}
Using $\sigma(0)=(45\pm 8)\ {\rm MeV}$ as reported in (\cite{GLS91}), 
we obtain
\begin{equation}
c_1^r=-0.99\, \pm\, 0.11\ {\rm GeV}^{-1}.
\label{c1rennum}
\end{equation}
While the second term in the brackets in (\ref{c1renfix}) amounts to 
$22.5\ {\rm MeV}$, the third term adds another $10\ {\rm MeV}$. Terms 
of order $M_\pi^3$ and $M_\pi^4/\Delta$ are of similar size, as
expected from our general discussion.

Using the value (\ref{c1rennum}) we can also fix the 
renormalized nucleon mass in the chiral limit from (\ref{mN}) as
\begin{equation}
m^r=880\,\pm\,10\ {\rm MeV}.
\end{equation}
This quantity is rather stable and changes by only $\approx -5 {\rm MeV}$
compared to the order $p^3$ result.

Finally, we turn to the shift of the scalar form factor between the Cheng-
Dashen point and zero. This quantity is peculiar in the sense that to 
order $p^3$ in HBChPT, the one-loop graph in Fig. 2 b) (with delta
propagator replaced by a heavy nucleon propagator), is the only
contribution and yields the finite result
\begin{equation}
\sigma(2 M_\pi^2)-\sigma(0)=
{3 \dot{g}_A^2 \over 64 \pi F^2}\, M_\pi^3 \approx 7.5\ {\rm MeV} .
\label{sigmashiftp3}
\end{equation}
This is off by a factor of 2 from the empirical value of 15 MeV obtained 
from a dispersive analysis. (\cite{GLS91}) At next-to-leading order,
${\cal O}(p^4)$, the analysis has been performed in chiral SU(3). 
(\cite{BKM93,BM96}). Because the relation between coupling constants in 
SU(2) and SU(3) HBChPT, respectively, is not yet known, it is
difficult to asses the consequences of these analyses for the present
investigation. Also, in (\cite{BM96}) resonance saturation was used in
a hybrid version, i.e. by keeping relativistic delta resonances in
loops -- I admittedly do not understand the logic behind this approach. 

In the small scale expansion, the leading order effect of intermediate
deltas can be calculated without introducing any new parameter.
(\cite{HHK97c}) 
It can easily be shown that the loop diagram of Fig. 2 b) is the only
new contribution at order $\epsilon^3$. The result for 
$\sigma(2 M_\pi^2)-\sigma(0)$ was given first in
\cite{BKM93}, but in that paper it could not be identified as the leading
term in a systematic expansion. Using this result, and carefully
extracting first the infrared singular pieces in $M_\pi$, we obtain
the chiral expansion of Fig. 2 b) (\cite{HHK97c})
\begin{equation}
\sigma(2 M_\pi^2)-\sigma(0)|_{\Delta}=
{g_{\pi N\Delta}^2 \over 6 \pi^2 F^2}\ {M_\pi^4 \over \Delta}
\left[ {5\over 18}-{\pi \over 24}+
{5\over 6} \ln {2\Delta \over M_\pi} 
+{\cal O}\left( {M_\pi^2 \over \Delta^2}\right)
\right] 
\label{sigmashiftexp}
\end{equation}
Numerically, the delta contribution depends much on the 
$\pi N\Delta$-coupling constant employed. 
Using $g_{\pi N\Delta}=1.05 \pm 0.02$, we find
$\sigma(2 M_\pi^2)-\sigma(0)|_{\Delta}\approx 4.0\ {\rm MeV}$ for the 
full one-loop contribution, and 3.7 MeV if we take the leading term
in the expansion in $M_\pi$ as given in (\ref{sigmashiftexp}). 
The numerically dominant contribution arises from the chiral logarithm in 
Eq. \ref{sigmashiftexp}. Note that the scale
of the logarithm is fixed and given by $2 \Delta$. The coefficient of 
this logarithm is of order unity, and the overall scale is given by the 
strong coupling constant $g_{\pi N\Delta}$. This, together with the 
observation that $\Delta\approx 2 M_\pi$ provides a natural 
explanation of the numerically large correction.

To summarize, we have to leading order in the small scale expansion
\begin{equation}
\sigma(2 M_\pi^2)-\sigma(0)|_{\epsilon^3}\approx 11.5\ {\rm MeV}.
\end{equation}
The delta resonance therefore gives a large correction to the $p^3$
result, and it goes into the right direction. This is consistent
with what we know from the dispersive analysis, where the relevant 
absorptive part picks up a similar contribtuion from the delta region.
The remaining piece is attributed to further continuum contributions,
which will appear as higher order corrections in the chiral
expansion. The small scale expansion improves the ``convergence'' of the
perturbative series by moving important effects due to the delta
resonances to lower orders in the expansion. However, before any firm 
conclusion can be drawn, it is mandatory to consider the 
next-to-leading order effects, i.e. an order $\epsilon^4$ calculation
is clearly called for. 

\subsection{Resonance saturation}

We have seen that the delta resonances manifest themselves in
threshold processes of the $\pi N$-system in two distinct manners
\begin{enumerate}
\item
The heavy components of the delta resonance, $G_v^\mu$, contribute to 
local counterterms of the theory we called small scale expansion. These
contributions arise via exchange graphs and technically appear as
$1/m$-corrections when integrating out the heavy degrees of freedom
$G_v^\mu$, see section 2.3. An explicit example of this effect is
provided by the coupling of pions, nucleon and delta. The relativistic
effective lagrangian reads
\begin{equation}
{\cal L}_{\pi N\Delta}^{\rm relativ.}=
g_{\pi N\Delta} \bar \psi_\mu^a \left[ g^{\mu\nu}-(Z+{1\over2}) 
\gamma^\mu \gamma^\nu \right] u_\nu^a N
\label{L2rel}
\end{equation}
where $Z$ denotes a so called off-shell parameter. Performing the
$1/m$-expansion we obtain a local term of order $p^2$ contributing to 
$\pi N$-scattering
\begin{equation}
{\cal L}_{\pi N}^{(2)}={g_{\pi N\Delta}^2\over 2 m} {4\over 3}
(1+8 Z+12 Z^2) \bar H_v \left[ (S\cdot u)^2-{3\over 2} {\rm tr} 
(S\cdot u)^2 \right] H_v
\label{L2local}
\end{equation}
Note that these contributions scale like $1/m$. 

\item
The light components of the delta spin 3/2 field, $T_v^\mu$, are kept
in the theory and contribute via tree and loop graphs. Examples of
this sort have been given in the discussion of the scalar sector of
the $\pi N$-system. The diagrams which have to be calculated to a 
given order in the small scale expansion can be found by 
power counting in $\epsilon$. These contributions can be expanded in 
$M_\pi/\Delta$ and therefore in general scale like $(M_\pi/\Delta)^n$. 
\end{enumerate}
The discussion of the scalar sector given above suggests a
reformulation of resonance saturation for counterterms of HBChPT. In a
first step one calculates the amplitude of an arbitrary process to a
given order in $\epsilon$. Then, expanding around the limit 
$\hat m \rightarrow 0$, $\Delta$ fixed, the amplitude can be matched 
onto the corresponding HBChPT amplitude. The effects of both,
heavy and light components of the delta are thereby absorbed in the 
coupling constants of HBChPT. Note that chiral logarithms of the type 
$\ln (M_\pi/2 \Delta)$, appearing e.g. in Eq. (\ref{Rexp}), can be 
absorbed by the ${\cal O}(p^2)$ coupling constants showing up via
one-loop graphs at order $p^4$ in the chiral expansion. 
The main difference to the way baryonic spin 3/2 resonances are
conventionally treated, i.e. as pole exchange graphs of relativistic
spin 3/2 fields, is the ability to include the ``light'' degrees
of freedom of the delta in loop graphs and therefore to resum these
effects. Also, corrections to this procedure are controlled by the
small parameter $\epsilon/\Lambda$, with $\Lambda\in \{ 4\pi F_\pi,
m_N\}$. Since the expansion is systematic, this offers in principle 
the possibility to quantitatively controll the accuracy of the
approach. The question of how well this works is the subject of 
present investigations, and we hope to come back to it soon.

\section{Application: polarizabilities of the nucleon}

The technique described in section 2 has recently been applied to the
problem of nucleon Compton scattering and the polarizabilities of the
nucleon. (\cite{HHK97b}) At order $\epsilon^3$ large effects due to 
$\Delta(1232)$ to the electric, $\alpha_E$, magnetic, $\beta_M$, and 
spin polarizability, $\gamma$, have been found. Whereas the results
for $\beta_M$ and $\gamma$ were expected on the basis of previous
analyses, the large effect on $\alpha$ came rather as a surprise. I
refer the reader to the talk by Holstein for a review of the subject
as well as for basic definitions. (\cite{Holstein})
Here I would like to provide a further example to show how large effects
entering at higher orders in the chiral expansion can show up at
leading order in the $\epsilon$-expansion. 

The ${\cal O}(\epsilon^3)$ results are
\begin{eqnarray}
{\alpha}_E & = & \frac{e^2}{4\pi} \; \frac{1}{6\pi 
                               F_{\pi}^2} \; \frac{1}{M_\pi}\left\{ 
                               \frac{5 g_{A}^2}{16} + \frac{g_{\pi 
                               N\Delta}^2}{\pi} \frac{M_\pi}{\Delta} \left[
                               1 + \frac{1}{9} \log \left( \frac{2\Delta}{
                               M_\pi} \right) \right]\right\} \\
 & \approx & \left[12.2 \; \mbox{(N-loop)} + \; 0 \; \mbox{(delta-pole)} + 
             \; 4.2 \; \mbox{(delta-loop)} \right] \; \times 10^{-4} \; 
             {\rm fm}^3   \nonumber \\
{\beta}_M  & = & \frac{e^2}{4\pi} \; \frac{1}{6\pi 
                               F_{\pi}^2} \; \frac{1}{M_\pi} \left\{ 
                               \frac{g_{A}^2}{32} + \frac{M_\pi}{\Delta} 
                               \left[ \frac{b_{1}^2}{3\pi} \frac{(4\pi 
                               F_\pi)^2}{m_{N}^2} + \frac{g_{\pi N\Delta}^2}{
                               9\pi} \log \left( \frac{2\Delta}{M_\pi} 
                               \right) \right] \right\} \\
  &\approx & \left[1.2 \; \mbox{(N-loop)} + \; 7.2 \; \mbox{(delta-pole)} + 
             \; 0.7 \; \mbox{(delta-loop)} \right] \; \times 10^{-4} \; 
             {\rm fm}^3 \nonumber 
\label{alphabeta}
\end{eqnarray}
and 
\begin{eqnarray}
\gamma &=& \frac{e^2}{4\pi} \; \frac{1}{216\pi^2 \; F_{\pi}^2} \;
\frac{1}{M_{\pi}^2} \left[ 9 g_{A}^2 - 6 b_{1}^2 \; \frac{(4\pi F_\pi)^2}{
m_{N}^2} \; \frac{M_{\pi}^2}{\Delta^2} - 4 g_{\pi N\Delta}^2 \; 
\frac{M_{\pi}^2}{\Delta^2} + \; \right]  \\
              &\approx & (4.6\mbox{ (N-loop)}-2.4\mbox{(delta-pole)}-0.2
                         \mbox{(delta-loop)})\times 10^{-4} \; {\rm fm}^4
                         \; . \nonumber
\label{gamma}
\end{eqnarray}
$b_1$ is a $\gamma N\Delta$-coupling constant entering at order $p^2$.
For simplicity we have expanded the full ${\cal O}(\epsilon^3)$
expressions to leading order in $M_\pi/\Delta$ which numerically
yields already the bulk part of the effect. However, the numbers 
given correspond to the full ${\cal O}(\epsilon^3)$ contributions, and
we used coupling constants determined from delta decay widths to
leading order in the small scale expansion. (\cite{HHKK97})

We observe that for $\alpha_E$ and $\beta_M$ the delta contributions
are suppressed by one chiral power, $M_\pi$, with respect to the
${\cal O}(p^3)$ $\pi N$-loop contribution. Numerically they are
nevertheless important, and this can be related to the appearance of
the small denominator $\Delta$. As to the spin polarizability
$\gamma$, the $\Delta$-pole contribution is large and opposite in sign
to the ${\cal O}(p^3)$ $\pi N$-loop contribution; it goes in the right
direction to bring about agreement with the positive sum rule value.
Formally, this effect is suppressed by two powers of $M_\pi$ with
respect to the leading order piece as obtained in HBChPT. It is
therefore of order $p^5$ and will show up only as part of a two-loop
calculation in HBChPT. The spin polarizability of the nucleon is a 
prime example of how a theory with delta degrees of freedom can
include important effects or higher orders in the chiral expansion 
already at leading order of the perturbation series. 

Although the effects discussed are encouraging and important, they
are by itself of course not sufficient to draw any firm conclusion. 
For instance, it is known from a $p^4$ HBChPT calculation
(\cite{BKMS94}) that the large and positive delta contribution to
$\beta_M$ is to a large extent cancelled by loop graphs including 
$p^2$-vertices. In order to see such effects as well as other
important corrections coming in only at next-to-leading order, an order
$\epsilon^4$ calculation is absolutely necessary.

\section{Conclusions}

Heavy Baryon Chiral Perturbation Theory is a powerful tool to study the 
interactions of pions, nucleons and photons in the low energy regime. 
The spin 3/2 delta resonances influence the effective low energy 
theory substantially, due to both the small mass difference 
$\Delta=m_\Delta-m_N$ and the large $\pi N\Delta$-coupling constant. 
HBChPT including the $\Delta(1232)$ as an effective degree of freedom
has been formulated recently, and it is shown that the theory admits 
a systematic expansion in the small scales $M_\pi$, soft momenta and 
the mass difference $\Delta$, collectively denoted by $\epsilon$. The
relation of this theory to HBChPT has been discussed on the example
of the scalar sector of the $\pi N$-system. In particular the
different meaning the counterterm couplings play in the two theories has
been stressed. The limit $\hat m\rightarrow 0$, $\Delta$ fixed can be
used to calculate the effects of $\Delta(1232)$ on the counterterm
coupling constants of HBChPT. The method includes all effects of the
delta, $1/m$ suppressed terms due to exchange of the heavy components as
well as terms scaling with $1/\Delta$ due to tree and loop graphs 
involving light components of the delta. Corrections to this procedure
are controlled by the parameter $\epsilon/\Lambda$, witch 
$\Lambda\in \{4 \pi F_\pi, m_N\}$. Phenomenologically, including 
$\Delta(1232)$ as an effective degree of freedom has the advantage of
moving large effects due to the delta resonance to lower order in the
expansion, thereby improving the ``convergence'' of the perturbation
series. The shift of the scalar form factor of the nucleon, 
$\sigma(2 M_\pi^2)-\sigma(0)$, as well as the spin polarizability 
$\gamma$ are examples of this sort discussed in this talk. As to the
electric and magnetic polarizabilities $\alpha$ and $\beta$,
respectively, large delta effects at ${\cal O}(\epsilon^3)$ seem to
spoil the previously found agreement between theory and experiment. 
Clearly, the next-to-leading order corrections of order $\epsilon^4$
have to be worked out, and this is one of the main directions for
future work in this framework. It will also be necessary to adress
other processes like $\pi N$-scattering, where abundant data will help
to determine unknown coupling constants. We have not discussed in this
talk applications to processes with excitation energies in the delta
region, like the E2/M1 multipole ratio measured at MAMI. For such 
applications it is necessary to have a theory which includes the
delta resonance explicitly. The formalism presented in section 2
is well suited to deal with this situation, and work in this direction
is well under way.

\noindent
{\bf Acknowledgments}

I would like to thank T. Hemmert and B. Holstein for an enjoyable
collaboration and the organizers for the interesting and stimulating 
workshop.

\end{document}